# Surface tension effects on flow dynamics and alveolar mechanics in the acinar region of human lung

Isabella Francis[1], Suvash C. Saha[1, *]

[1]School of Mechanical and Mechatronic Engineering, Faculty of Engineering and Information Technology, University of Technology Sydney, NSW, Australia

*Corresponding author: Suvash C. Saha; Email: Suvash.Saha@uts.edu.au

**Abstract.** Computational fluid dynamics (CFD) simulations, in-vitro setups, and experimental ex-vivo approaches have been applied to numerous alveolar geometries over the past years. They aimed to study and examine airflow patterns, particle transport, particle propagation depth, particle residence times, and particle-alveolar wall deposition fractions. These studies are imperative to both pharmaceutical and toxicological studies, especially nowadays with the escalation of the menacing COVID-19 virus. However, most of these studies ignored the surfactant layer that covers the alveoli and the effect of the air-surfactant surface tension on flow dynamics and air-alveolar surface mechanics. The present study employs a realistic human breathing profile of 4.75s for one complete breathing cycle to emphasize the importance of the surfactant layer by numerically comparing airflow phenomena between a surfactant-enriched and surfactant-deficient model. The acinar model exhibits physiologically accurate alveolar and duct dimensions extending from lung generations 18 to 23. Airflow patterns in the surfactant-enriched model support previous findings that the recirculation of the flow is affected by its propagation depth. Proximal lung generations experience dominant recirculating flow while farther generations in the distal alveolar region exhibit dominant radial flows. In the surfactant-enriched model, surface tension values alternate during inhalation and exhalation, with values increasing to 25 mN/m at the end of inhalation and decreasing to 1 mN/m at the end of exhalation. In the surfactant-deficient model, only water coats the alveolar walls with a high surface tension value of 70 mN/m. Results showed that surfactant deficiency in the alveoli adversely alters airflow behavior and generates unsteady chaotic breathing through the production of vorticities, accompanied by higher vorticity and velocity magnitudes. In addition, high air-water surface tension in the surfactant-deficient case was found to induce higher shear stress values on the alveolar walls than that of the surfactant-enriched case. Overall, it was concluded that the presence of the surfactant improves respiratory mechanics and allows for smooth breathing and normal respiration.

## 1. Introduction:

Human lungs are vital organs of the respiratory system, responsible for oxygen and carbon dioxide exchange through the capillary networks of tiny air sacs in the distant lungs, known as the alveoli. Type II alveolar cells synthesize, store, and release surfactant lipids and proteins in the lungs. This surfactant resides on top of a thin water layer that covers the inner surface of the alveoli. Intrinsically, surfactant production of babies begins around week 24 of gestation and increases rapidly in weeks 34 and 35 (8 months pregnancy). Hence, pre-born babies suffer from surfactant deficiency and develop a respiratory distress syndrome called Hyaline Membrane Disease (HMD). Surfactant deficiency is also observed in adults who suffer from respiratory diseases such as acute respiratory distress syndrome (ARDS) (Nkadi et al., 2009). In the absence of pulmonary surfactant, an interaction between air and water molecules occurs inside the alveoli during inhalation. Dissimilar air and water molecules tend to pull farther from each other, resulting in high surface tension. The water molecules pull each other toward the alveolar surface and create a thinner water layer. As a response, the alveoli start collapsing after 48 to 72 hours, and damaged cells known as "hyaline membranes" accumulate in the airways, causing difficulty breathing accompanied by rattling and bubbling sounds. Thus, the existence of the pulmonary surfactant in the alveoli is crucial to reducing the surface tension of the air-liquid interface by pulling the water molecules back upwards. As a result, the water layer returns to its natural thickness, and the air-water surface tension decreases.

An acinus is a group of alveoli and alveolated ducts located distal to a single terminal bronchus beyond the fifteenth generation of the lung (Haefeli-Bleuer and Weibel, 1988). Many computational studies have focused on local alveolar fluid flow characteristics in multiple acinar generations represented by the alveolar to ductal flow rate $Q_A/Q_D$ (Sznitman et al., 2007) and dimensionless numbers such as the Reynolds, Strouhal (Tsuda et al., 2008), and Womersley numbers (Sznitman et al., 2009). Later studies shifted their focus onto the effect of the gravitational and convection aerodynamic force on the transport and deposition of micron-sized particles in the acinus region for healthy alveoli (Ma and Darquenne, 2011, Koullapis et al., 2018) and diseased (emphysematous) alveoli along with accompanied unsteady flow visualizations (Xi et al., 2021). Some studies have included the effect of Brownian motion to examine nanoparticle transport and deposition in different scenarios. For example, Xi and

Talaat (2019) examined the influence of alveolar wall motion and variations in the size of interalveolar septal apertures. A further study considered the effect of different alveolar sizes on 10, 50, 200, and 800-nm nanoparticle deposition in terminal alveolar sacs (Xi et al., 2020). In-vitro studies (Fishler et al., 2017, Dong et al., 2021) and experimental studies using Particle Image Velocimetry (Oakes et al., 2010, Sznitman, 2013) have also been popular in observing fluid flow characteristics and particle dispersion across acinar trees. The abovementioned studies adopted idealised sinusoidal breathing profiles for the inhalation and exhalation phases instead of a realistic, piecewise, high-order polynomial profile. These have also neglected the presence of the surfactant on the alveolar surfaces.

One recent study applied constant surface tension values on the surfaces of spherically shaped alveoli and observed the variations in the Reynolds number, Womersley number, and ductal to alveolar flow rate ratios (Dong et al., 2020). Another study performed finite element analysis on an acinus geometry segmented from computed tomography (CT) images to examine the increase in local alveolar anisotropic deformation due to high surface tension (Koshiyama et al., 2019). Other studies included the surface tension effect in different ways using fluid-structure interaction (FSI) on simplified alveolar models to represent the movement of the alveoli during inhalation and exhalation. Chen et al. (2021) employed fluid-structure interaction (FSI) on a three-dimensional honeycomb-like alveolar model. They investigated the effect of both pulmonary fibrosis and surface tension on alveolar mechanics (pressure drop, displacement, velocity, and von Mises stress) in a disease state known as Diffuse Alveolar Damage (DAD). The disease state was reflected by increasing alveolar tissue thickness assuming a nonlinear hyper-elastic material and by decreasing surfactant concentrations. They accounted for surface tension by using a dynamic compression-relaxation model, which determines the rate of surface tension change based on the surface area and parameters extracted from experimental data. Results revealed that high surface tension values contributed to hysteresis of lung tissue with minor changes in airflow rate, volume-pressure relationship, and alveolar resistance. Monjezi and Saidi (2016) incorporated surface tension into a honeycomb-like alveolar geometry to assess its ramifications on alveolar deformations (stretches) and stresses. They adjusted the Mooney Rivlin model's parameter constants to capture the surface tension effect. The aforementioned studies ignored the multiple acinar generations and mainly focused on alveolar tissue mechanics for alveolar sacs. These have also neglected to address airflow changes with surface tension variations.

The current work realistically represents the variations in the airflow characteristics and air-alveolar surface mechanics between a surfactant-deficient and a surfactant-enriched acinar model. The study aims to demonstrate the complications resulting from the lack of pulmonary surfactant in the few days before the collapse of the alveoli.

## 2. Methods:

### 2.1. Acinar model:

The idealised alveolar model (Fig. 1) consists of a single path airway extending from lung generation 18 to generation 23 by tracing one duct from each bifurcation. The geometry in Fig. 1 starts with two respiratory bronchioles in lung generation 18 (Gen 18) and bifurcates till reaching the alveolar sacs in generation 23 (Gen 23). Inner bifurcating angles of 30 degrees are employed as approximated in the lungs (Sauret et al., 2002). The duct lengths and diameters were obtained from Sznitman (2013) and Haefeli-Bleuer and Weibel (1988), with duct diameters and lengths decreasing as generation number increases. Specifically, duct lengths decrease from 765 µm in generation 18 to 575 µm in generation 23. Likewise, duct diameters decrease gradually from 330 µm to 240 µm.

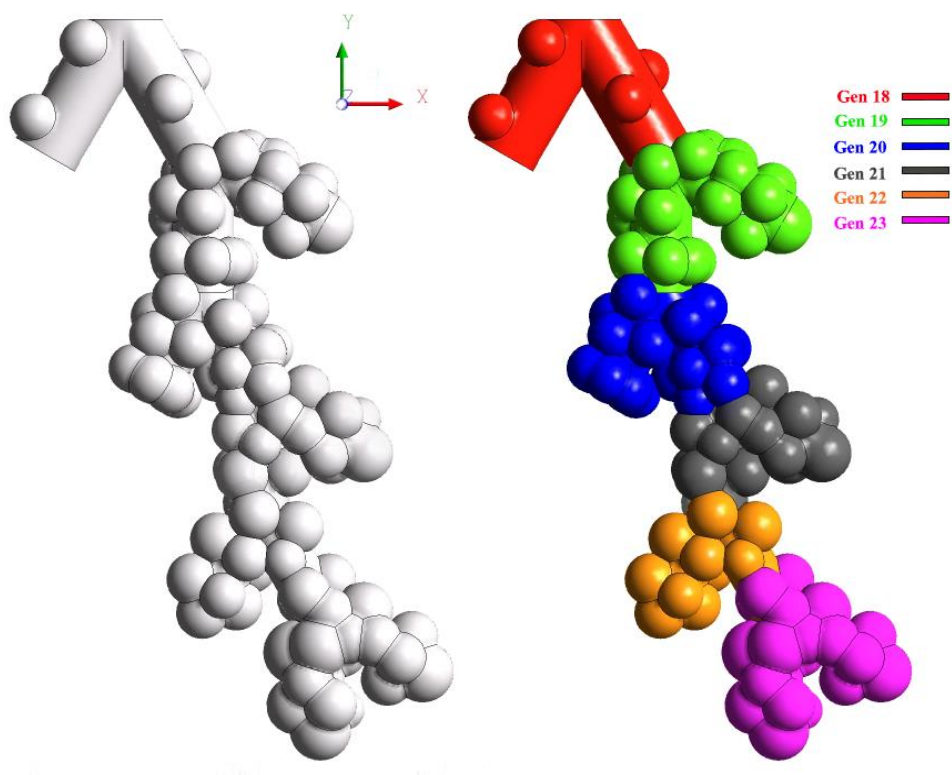

Fig. 1: Acinar model with different generation numbers

Alveoli are idealised as hemispheres. The total alveoli number and the size of each alveolus match the actual in-vivo lung measurements. For the latter, studies have shown that the mean volume of a single alveolus is within the range of $[3.3 \rightarrow 4.8] \times 10^6$ µm$^3$ irrespective of lung size (Ochs et al., 2004). Since the volume of a hemisphere is directly proportional to its radius where $Volume = 2\pi r^3/3$, then the range of the alveoli radii is $[0.11636 \rightarrow 0.13184]$ mm. As for the total number of alveoli, it has been established that one cubic millimeter (1mm$^3$) of lung parenchyma contains around 170 alveoli (Ochs et al., 2004). The volume of the acinar model in Fig. 1 ($Volume = 0.89629 \; mm^3$) gives rise to about 150 alveoli. These are primarily distributed between Gen 19 to Gen 23 as exhibited in the lungs (Patwa and Shah, 2015).

## 2.2. Realistic Breathing Velocity Profile:

As presented by Russo and Khalifa (2011), a realistic breathing velocity profile includes an exhalation period of 2s, a pause of 1s, and an inhalation period of 1.75s, providing a total breathing cycle time of 4.75s. The profile corresponds to a volume flow rate of 6 litres per minute. The number of breaths per minute, the normal tidal volume, and the mass of the studied individual are calculated from the following equations:

$$\text{Total cycle time} = \frac{60 \text{ seconds}}{\text{Breaths per minute}} = 4.75\text{s} \quad (1)$$

$$\rightarrow \text{Breaths per minute} = 12.63 \cong 13$$

$$\text{Volume flow rate} = \text{Tidal volume} \times \text{Respiratory rate}$$

$$\rightarrow \text{Tidal volume} = \frac{\text{Volume flow rate}}{\text{Respirstory Rate}} = \frac{6\text{l/min}}{13 \text{ BPM}} = 460 \text{ ml in one breath} \quad (2)$$

Since the normal tidal volume is approximately 7 ml/kg, regardless of age, then the mass of the studied individual can be found from:

$$7\frac{\text{ml}}{\text{kg}} \times \text{mass (kg)} = 460\text{ml}$$

$$\rightarrow \text{mass} = \frac{460\text{ml}}{7\text{ml/kg}} = 65.7 \text{ kg}$$

Therefore, the studied case corresponds to a person weighing 65.7 kg with a respiratory rate of 13 breaths per minute and a tidal volume of 460 ml. Since every parent duct in the same lung generation is bifurcated into two daughter ducts, the volume flow rate of a particular lung generation can be found through the following equation (Kumar et al., 2009):

$$\dot{V}_{\text{generation number}} = \frac{\dot{V}_{\text{inlet}}}{2^{\text{generation number}}} \quad (3)$$

$$\dot{V}_{18} = \frac{6 \text{ l/min}}{2^{18}} = 2.28882 \times 10^{-5} \text{ l/min/}$$

The volume flow rate is related to the average velocity by the formula:

$$\dot{V} = \text{Area} \times \text{Average Velocity} \quad (4)$$

Where the diameter of the $18^{th}$ generation duct is $d = 330 \text{ μm}$,

$$\rightarrow \text{Area} = \frac{\pi}{4}d^2 = 8.55299 \times 10^{-8} \text{ m}^2$$

$$\rightarrow \text{Average\_velocity} = \frac{\dot{V}}{\text{Area}} = 0.00446 \frac{\text{m}}{\text{s}} \quad (5)$$

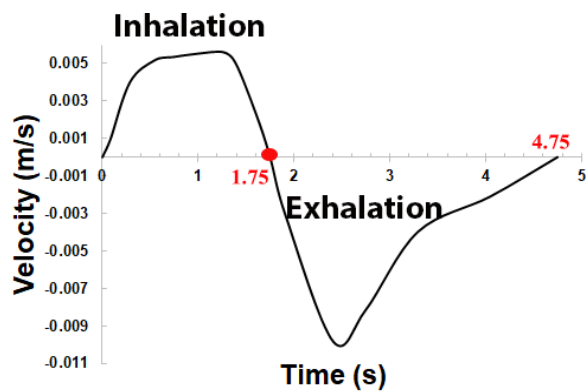

Fig. 2: Realistic breathing velocity profile for one complete breathing cycle

The realistic breathing velocity profile at the inlet of Gen 18 is obtained as shown in Fig. 2. The inhalation phase starts from time 0 to 1.75 seconds, exhibiting a maximum velocity of 0.005 m/s at around 1.3 seconds, decreasing to 0 at the end of inhalation. In the exhalation phase, velocity values are assigned a negative value to represent a change in the flow direction. Exhalation begins at 1.75 seconds and finishes at 3.75 seconds, followed by a 1-second pause. During exhalation, the absolute maximum velocity reaches 0.01 m/s at 2.4s and drops to 0 m/s at the end of the breathing cycle.

## 2.3 Mesh generation and mesh independent test:

In order to ensure accurate results, six mesh sizes were compared, from a coarse mesh with 60 μm tetrahedral volume elements to a finer mesh with 20 μm tetrahedral volume elements. Boundary layers were included throughout the acinar model to accurately evaluate the near-wall airflow behavior as depicted in Fig. 3(a). The volume-weighted averages of the airflow vorticity magnitude (s$^{-1}$) and the total airflow pressure (Pa) were compared for the various mesh element sizes as shown in Fig. 3(b). Oxygen and Carbon dioxide diffusion across alveolar walls in submillimeter acinar airspaces does not result in significant air density changes, rendering the flow effectively non-compressible (Davidson and Fitz-Gerald, 1972). Therefore, inspiratory airflow was assumed to be isothermal (37°C) and incompressible ($\rho =$ 1.139 $kg/m^3$). The inlet Reynolds number is less than unity even at peak velocity values (Karl et al., 2004), and thus the flow field was solved using the laminar model. Fig. 3(b) illustrates that the coarser the mesh, the greater the instability in the total pressure values and vorticity magnitudes. The variations of the two figures decreased at the mesh element size of 50 µm, with negligible variation (< 1%) when further refining the mesh to 20 µm. As a result, any computational mesh below 50 µm can be chosen, and thus 30 µm was adopted for all subsequent cases.

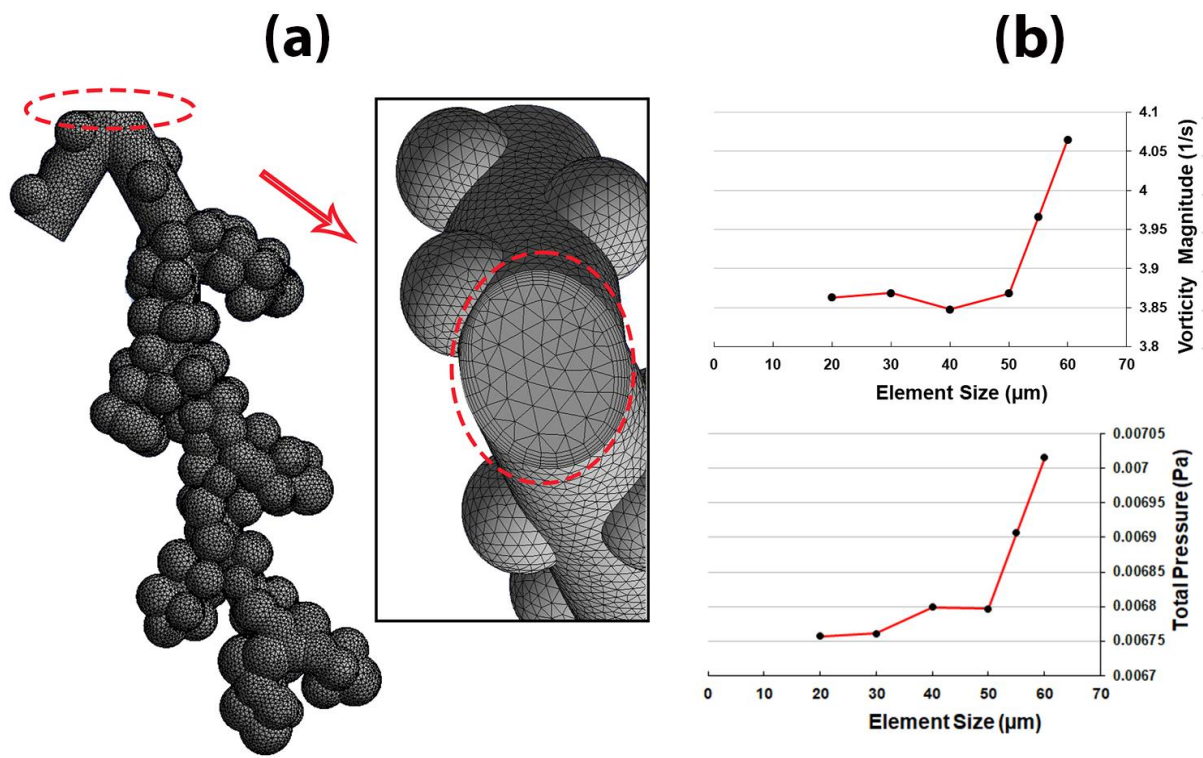

Figure 3: Computational mesh and sensitivity studies: (a) Computational mesh with a zoomed-in view of the tetrahedral elements and near-wall boundary layers, (b) Mesh-independent tests of the airflow vorticity magnitude (s$^{-1}$) and total pressure (Pa).

## 2.4 Addition of surfactant:

### 2.4.1. Surfactant Surface Tension

Pulmonary surfactant is a complex mixture of phospholipids (PL) and proteins (SP) that reduces the surface tension at the air-liquid interface inside the alveoli. For a surfactant layer to be applied to the inner surfaces of the alveolar model, its density and viscosity must be determined. A commonly used artificial surfactant (Survanta) that supplements the natural surfactant lacking in premature infants was adopted with a density $\rho = 25 \frac{kg}{m^3}$, and a dynamic viscosity $\mu = 3.25 \times 10^{-5} \frac{m^2}{s}$ (Lu et al., 2009). The air-surfactant surface tension varies during the inhalation and exhalation phases. During inspiration, the alveolus expands, resulting in a low surface concentration of surfactant and thus high surface tension values. At expiration, however, the alveolus contracts, and the surfactant concentration increases, thus reducing the surface tension to near-zero values. Fig. 4 demonstrates the variation in surface tension values during one breathing cycle. At the end of inhalation (1.75 seconds), the surface tension values peak at 25 mN/m and gradually decrease to around 0 mN/m at the end of exhalation (4.75 seconds) (Veldhuizen et al., 1998).

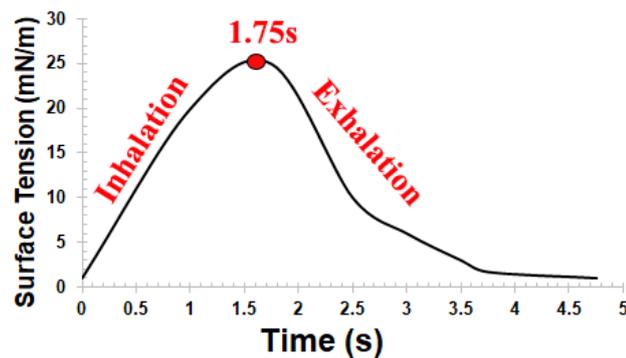

Figure 4: Variation of the air-surfactant surface tension values during inhalation and exhalation

## 3. Numerical methods:

Since air density changes are negligible within the acinus, airflow is considered unsteady and incompressible. Navier-Stokes equations (conservation of flow mass and momentum) for unsteady and incompressible flows were solved using a commercial finite-volume-based program, ANSYS Fluent 20.2 (Canonsburg, PA), to simulate the airflow motion. ANSYS MESHING (Ansys, Inc.) was utilised for computational mesh generation, and a user-defined function was applied to model the inhalation and exhalation velocity breathing profile. The multiphase Volume of Fluid (VOF) model was activated with an implicit formulation, implicit body force, and sharp interface modeling to distinguish air volume from surfactant or water. Surface tension force modeling was applied with a continuum surface stress model and a 0° surfactant or water wall adhesion. For the surfactant-enriched case, the primary phase was defined as air with density and dynamic viscosity properties at body temperature (37°C). The second phase was defined as surfactant with the properties shown in Table 1. For the surfactant-deficient case, the second phase was defined as liquid water with density and dynamic viscosity properties at body temperature. Water surface tension was set to 70 mN/m, but the surface tension of surfactant was set to vary, as depicted in Fig. 3. A pressure-velocity coupling scheme, PISO, with skewness and neighbor correction of 2, and a second-order body-force weighted pressure-based spatial discretization were adopted. In addition, a modified high-resolution interface capturing (HRIC) VOF spatial discretization was implemented to capture the sharp interface between air and surfactant or water by obtaining the face fluxes of all cells, including those adjacent to the interface. The surfactant layer spreads with a thickness of 0.1 µm on the inner surfaces of the alveoli (Siebert and Rugonyi, 2008). Therefore, phase 2 was patched with a 0.1 µm distance to the alveolar walls in both cases. For one complete breathing cycle of 4.75s, 0.001s of step time was used with 4750 steps per second.

### 3.1. VOF Model Equations:

- Continuity equation for primary phase 1 and secondary phase 2:

$$\frac{\partial \rho}{\partial t} + \nabla \cdot (\rho \vec{u}) = S \tag{6}$$

Where $\rho = \alpha_1 \rho_1 + \alpha_2 \rho_2$ is the total density with $\alpha$ and $\rho$ being the volume fraction and the density, respectively, $\vec{u} = \frac{1}{\rho}(\alpha_1 \rho_1 \vec{u_1} + \alpha_2 \rho_2 \vec{u_2})$ is the mixture velocity, and S is a source term.

Only one momentum equation is solved for both phases throughout the domain, where both share the resulting velocity field. A set of properties and variables are then assigned to each control volume based on the local value of the volume fraction.

- Momentum equation of mixture:

$$\frac{\partial (\rho \vec{u})}{\partial t} + \nabla \cdot (\rho \vec{u} \vec{u}) = -\nabla p + \nabla \left( \mu (\nabla \vec{u} + \nabla \vec{u}^T) \right) + \rho \vec{g} + \vec{T}_\sigma \tag{7}$$

$\vec{T}_\sigma$ is the surface tension force at the phase interface.

The VOF model assumes that the primary and the secondary phases are immiscible, meaning that in most computational cells, the volume fraction of each phase ($\alpha$) is either 0 or 1. However, at the interface between the two phases (air and surfactant or water in this case), the volume fraction is between 0 and 1 ($0 < \alpha < 1$), and the interface is tracked by solving the volume fraction equation (considered here for one or more secondary phases).

- Volume fraction equation for the secondary phases:

The interface tracking between the two phases is achieved by solving a continuity equation for the volume fraction of the secondary phases. This equation has the following form:

$$\frac{1}{\rho_q}\left[\frac{\partial}{\partial t}(\alpha_q \rho_q) + \nabla \cdot (\alpha_q \rho_q \vec{v}_q) = S_{\alpha_q} + \sum_{p=1}^{n}(\dot{m}_{pq} - \dot{m}_{qp})\right] \tag{8}$$

where $\dot{m}_{qp}$ is the mass transfer from phase q to phase p and $\dot{m}_{qp}$ is the mass transfer from phase p to phase q. By default, the source term on the right-hand side of the equation, $S_{\alpha_q}$, is zero, unless specified otherwise.

- The Implicit Scheme:

The implicit scheme computes the volume fraction values at the current time step, unlike the explicit method, which requires the volume fraction at the previous time step. The secondary phase volume fraction is iteratively calculated by solving a scalar transport equation at each time step.

$$\frac{\alpha_q^{n+1}\rho_q^{n+1} - \alpha_q^n \rho_q^n}{\Delta t} V + \sum_f (\rho_q^{n+1} U_f^{n+1} \alpha_{q,f}^{n+1}) = \left[S_{\alpha_q} + \sum_{p=1}^{n}(\dot{m}_{pq} - \dot{m}_{qp})\right] V \tag{9}$$

where $n + 1$ is an index for the current time step, $n$ is an index for previous time step, $\alpha_{q,f}^n$ is the face value of the $q^{th}$ volume fraction, computed from the second-order upwind, modified HRIC scheme, $V$ is the volume of a cell, and $U_f^n$ is the volume flux through the face based on normal velocity.

## 3.2. Weber number and Capillary number

The Weber number is a dimensionless quantity that determines the relative significance of a fluid's inertia to its surface tension. The Weber number is defined as $We = \rho L_c U^2/\sigma$ where $\rho$ and $U$ are the density ($kg/m^3$) and velocity ($m/s$) of air, respectively, $L_c$ is the characteristic length ($m$), and $\sigma$ is the surface tension between air and surfactant or water ($N/m$). The capillary number is a dimensionless quantity describing the interaction between viscous drag forces and surface tension forces acting across gas and liquid interface. The capillary number is defined as $Ca = \mu U/\sigma$ where $\mu$ and $U$ are the dynamic viscosity ($kg/m.s$) and velocity ($m/s$) of air, respectively, and $\sigma$ is the interfacial surface tension ($N/m$). Proximal lung generations experiencing laminar-to-turbulent transitional flows or fully developed turbulent flows with Reynolds numbers greater than unity ($Re \gg 1$) require the calculation of the Weber number to determine the impact of surface tension effects. Since the Weber number measures the ratio of the aerodynamic force to the surface tension force, a Weber number smaller than unity ($We \ll 1$) denotes that the surface tension energy dominates the kinetic energy. However, since the laminar airflow inside the lung acinus at distal lung generations is characterized by Reynolds numbers less than unity, $Re \ll 1$, the capillary number with $Ca \ll 1$ better represents the significance of the surface tension effects. Fig. 5 shows the variation of the capillary number for each surfactant case and water case. For the calculation of the capillary number, the air dynamic viscosity is taken to be $\mu = 1.927 \times 10^{-5}\ kg/m.s$ at normal body temperature (37°C), the surface tension value of water is $\sigma = 70\ mN/m$, that of surfactant is the same shown in Fig. 4, and the velocity variation with time is the same shown in Fig. 2, with absolute values taken for the exhalation phase.

Figure 5 illustrates that the Capillary number fluctuates between 0 and $2 \times 10^{-5}$ for the surfactant case and between 0 and $2.5 \times 10^{-6}$ for the water case throughout the breathing cycle. For the surfactant case, higher values of Ca are observed, with a peak occurring during exhalation at around 4.2s. For the water case, the maximum Ca value also occurs during exhalation at around 2.5s. A maximum percentage variation of 98.75% occurs between the two cases at 4s, where the surfactant shows significantly larger Ca values due to its lower surface tension values, especially during exhalation. The Capillary number exhibits gradually increasing then gradually decreasing values from 0 to 1.75s during inhalation, following a pattern similar to the velocity profile. The results of both cases with $Ca \ll 1$ show that the surface tension force is substantially larger than the viscous force, thus proving the importance of considering its effects on the surrounding airflow and alveolar surface.

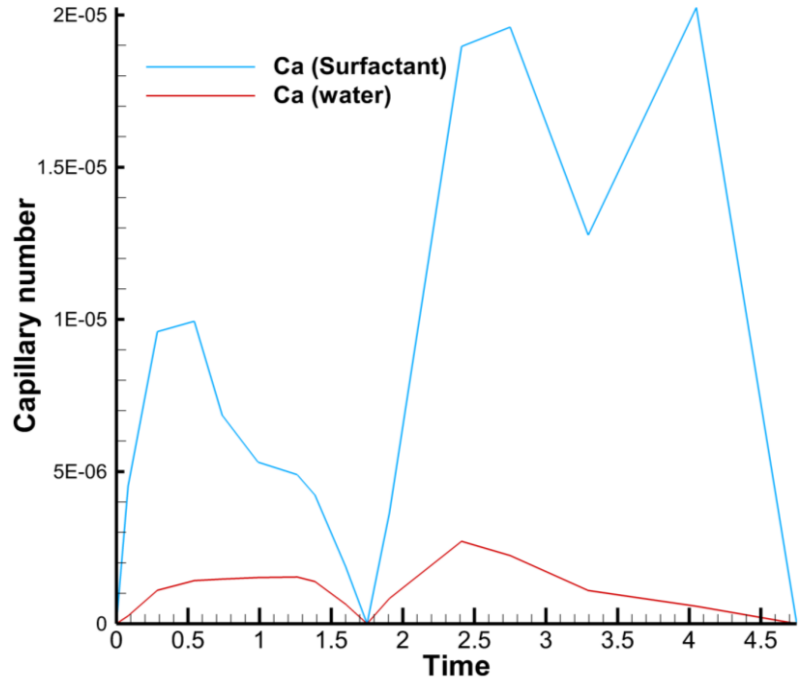

Fig. 5: Capillary number, $Ca = \frac{\mu U}{\sigma}$, variation throughout the breathing cycle for the surfactant case and the water case

## 4. Results and discussion:

### 4.1. Flow behavior

Figures 6A and 6B depict a comparison of volume-weighted average velocity magnitudes and average vorticity magnitudes respectively between a surfactant-enriched model (low surface tension) and a surfactant-deficient model (high surface tension). Since vorticity is a measure of the rotational velocity of fluid molecules, both velocity and vorticity magnitudes follow the same pattern. In Fig. 6A, the surfactant-enriched model with low surface tension has reduced values of average velocity magnitudes throughout the breathing cycle. In the inhalation phase of 1.75 seconds, a maximum variation of 8.69% occurs between the two cases at around 0.875s with a velocity value of 0.00115 m/s for the low surface tension case and 0.00125 m/s for the high surface tension case. During the exhalation phase of 3s (including pause), the difference between the two cases is more evident, especially between 2.5s and 3.7s. A maximum variation of 11.9% is observed at 2.7s with 0.0021 m/s for the low surface tension case and 0.0023 m/s for the high surface tension case. Figure 6B demonstrates that the surfactant-enriched model with low surface tension exhibits lower values for average vorticity magnitudes throughout the respiratory cycle. For both inhalation and exhalation, variations between the two cases are evident. During the inhalation phase, a maximum variation of 20% occurs in the middle of the phase with a 17.5 (s⁻¹) vorticity magnitude for the low surface tension case versus a higher 21 (s⁻¹) vorticity magnitude for the high surface tension case. At 2.7s during the exhalation phase, a maximum variation of 40.625% occurs with 32 (s⁻¹) for the low surface tension case versus a much higher 45 (s⁻¹) for the high surface tension case. It can be concluded from Figs. 6A and 6B that high surface tension in the absence of a surfactant increases airflow velocity and vorticity magnitudes.

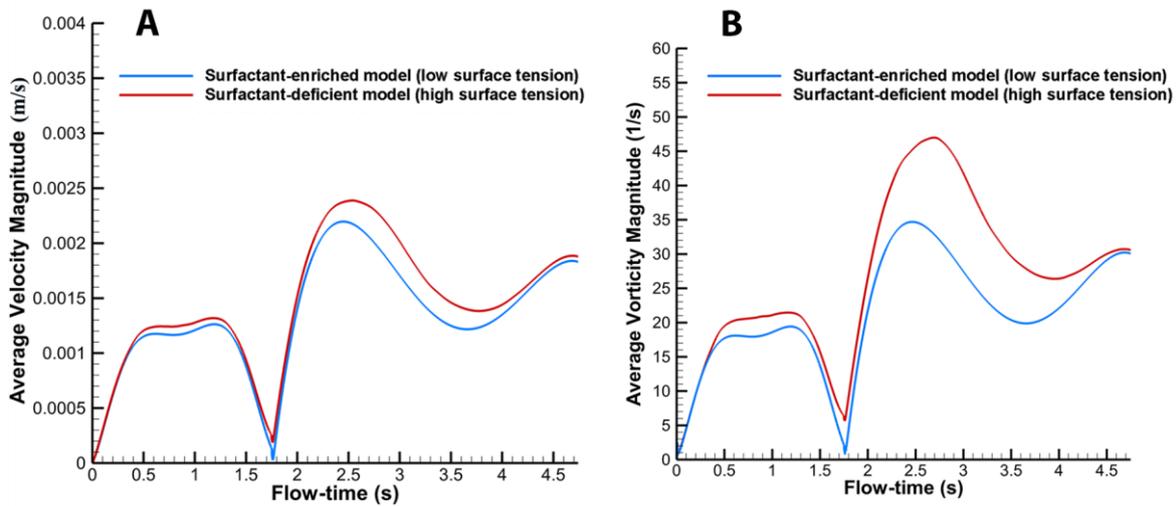

Figure 6: Comparison between a surfactant-enriched model and a surfactant-deficient model with volume-weighted averages of (A) Velocity magnitudes (m/s) and (B) Vorticity magnitudes (1/s)

One way to describe the effect of high surface tension inside an alveolus is by Newton's third law, which states that there is a reaction for every action. The action here is the force generated by the water molecules on the alveoli surface as they pull away from the surrounding air molecules, reducing the thickness of the water layer. The reaction is an equal force in the opposite direction, from the alveoli surface towards the inside of the alveoli cavity. This reaction force is the one that promotes the collapse of the alveoli in a surfactant-deficient lung where high surface tension forces exist. As a result, the pressure exerted by the alveoli will be equal to $P = \frac{F}{A}$, where $P$ is the collapsing pressure, $F$ is the force exerted by the alveolus on the water layer, and $A$ is the alveolus surface area.

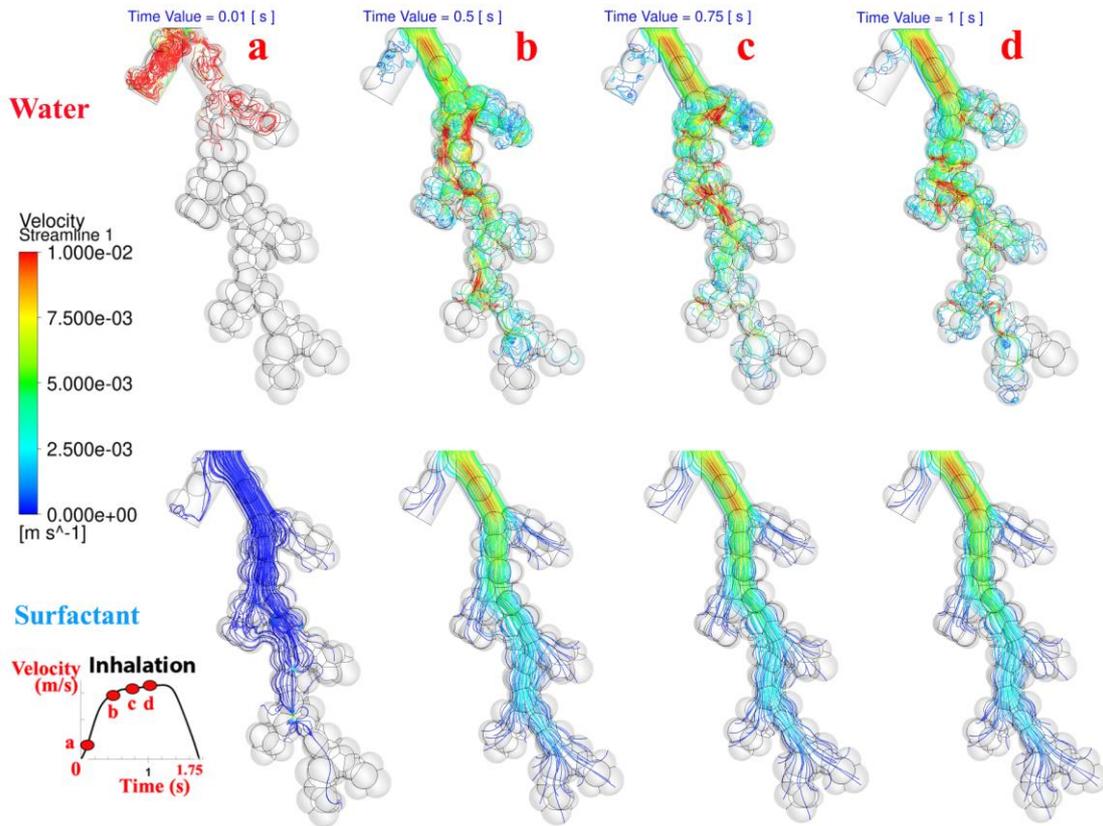

Figure 7: Visualization of different velocity streamline patterns for the surfactant and the water case during the first half of the inhalation phase

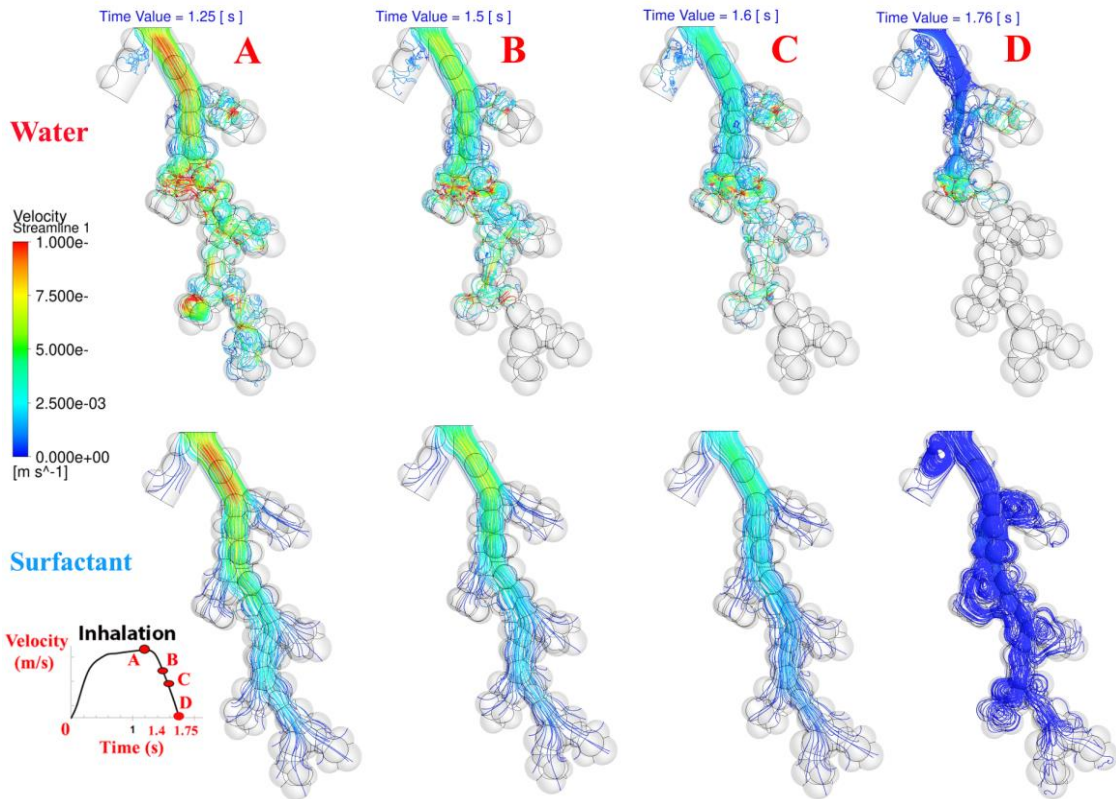

Figure 8: Display of contrasting velocity streamline patterns for the surfactant and the water case during the second half of the inhalation phase

Another way is by Laplace's law which provides a relationship between surface tension and pressure in spheres. In the case of alveoli, the air-liquid surface tension inside an alveolus is directly proportional to the inside air pressure through the following formula:

$$T = \frac{P \times R}{2} \quad (10)$$

Here $T$ is the air-liquid surface tension, $P$ is the pressure inside a spherically shaped bubble (alveolus), and $R$ is its radius. Thus, increased surface tension in the surfactant-deficient case increases the air pressure inside the alveoli. During inhalation, air flows from a higher-pressure region to a lower pressure region, drawing air into the alveoli. Therefore, the increased air pressure inside the alveoli due to high surface tension hinders smooth breathing. As a result, vortices are created, increasing both velocity and vorticity magnitudes. Figure 7 depicts this phenomenon, showing velocity streamlines during the first half of the inhalation period. For simplicity, the surfactant-deficient case is referred to as water, and the surfactant-enriched case is referred to as surfactant. These two cases exhibit significantly different velocity magnitudes and streamline patterns. At 0.01s, high-velocity values of 0.01 (m/s) are evident in the water case, with very low velocities of less than an order of magnitude in the surfactant case. At 0.5s, 0.75s, and 1s, smooth breathing can be observed in the surfactant case where streamlines are radially directed towards the alveoli, while chaotic breathing with recirculating streamlines is apparent in the water case. Similar phenomena are evident during the second half of inhalation, as shown in Fig. 8. The water case exhibits greater vorticity and velocity magnitudes at 1.25s, 1.5s, and 1.6s compared to the surfactant case, with decreasing velocities towards the end of inhalation. At 1.76s, the inhalation phase is completed, and exhalation begins. This explains the large vortices in the surfactant case and the half-full model in the water case.

An area-weighted average of vorticity magnitudes for the surfactant and water cases on each of the Gen 18 to Gen 23 surfaces is shown in Figure 9. In the surfactant case, Gen 18 experiences the highest vorticity values, reaching a maximum of 22 ($s^{-1}$) during inhalation and 47 ($s^{-1}$) during exhalation, while Gen 23 experiences only a maximum of 5 ($s^{-1}$). Gen 18 also has the highest vorticity values in the water case, reaching 90 ($s^{-1}$) at the end of exhalation, which is a 100% increase compared to the surfactant case. In all generations, the water case exhibits significantly higher surface vorticity values compared to the surfactant case. In addition, it can be seen from the surfactant case that as the generation number goes higher, the vorticity magnitude decreases. The reason is that as air flows through the acinus, more frequent contact occurs with the alveoli walls, thus reducing the recirculating flow and decreasing the vorticity magnitudes. This pattern is not the same for the water case where all the generations after Gen 18 have intersecting vorticity magnitudes throughout the breathing cycle.

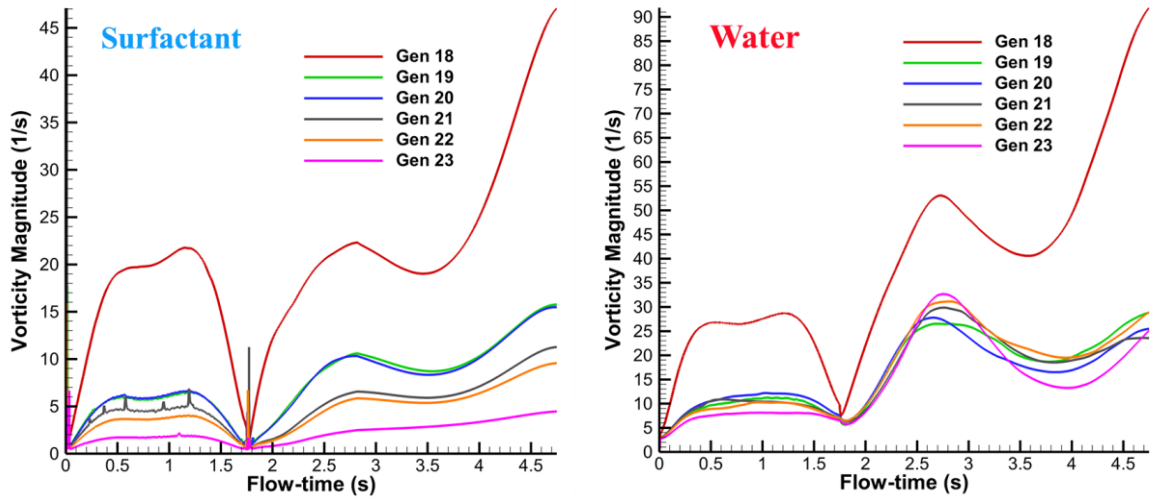

Figure 9: Area-weighted average of vorticity magnitude (1/s) on each generation surface for the surfactant case (left) and water case (right)

**4.2. Surface mechanics**

Figure 10 shows an area-weighted average of shear stress for the surfactant and water cases on the surfaces of Gen 18 to Gen 23. For the surfactant case, shear stress decreases as the generation number increases. This is due to the increased air-alveolar contact that reduces the erratic unsteady behavior of the flow in higher generations and thus creates less alveolar shear stresses. Gen 18 reaches a maximum value of 0.00275 Pa during the inhalation phase and a peak value of 0.0041 Pa during the exhalation phase. Gen 23 has the lowest shear stress values with a minimum of 0.0005 Pa. The pattern differs for the water case and much higher shear stress values are found for all generation numbers. The reason is that the high surface tension of the air-water interface induces a more turbulent flow with hectic vortices and larger velocity gradients that are directly associated with the shear stress force on the alveolar walls. Gen 18 for the water case displays the highest shear stress values, ranging from 0.024 Pa at the start of inhalation to 0.001 Pa at the end of the breathing cycle. The reason behind the decreasing shear stress values for all generations as time increases is that the volume fraction of water covering the alveolar walls is set to decrease with time. This is to mimic the *in-vivo* surfactant-deficient acinus where water might only partially cover the alveolar surfaces. However, even with the decrease in the water volume fraction and thus the shear stress, the lowest value reached by all generations in the water case (0.001 Pa) at the end of exhalation remains higher than that found in the surfactant case (0.0005 Pa). This highlights the immense contribution of water surface tension in increasing the surface shear stress.

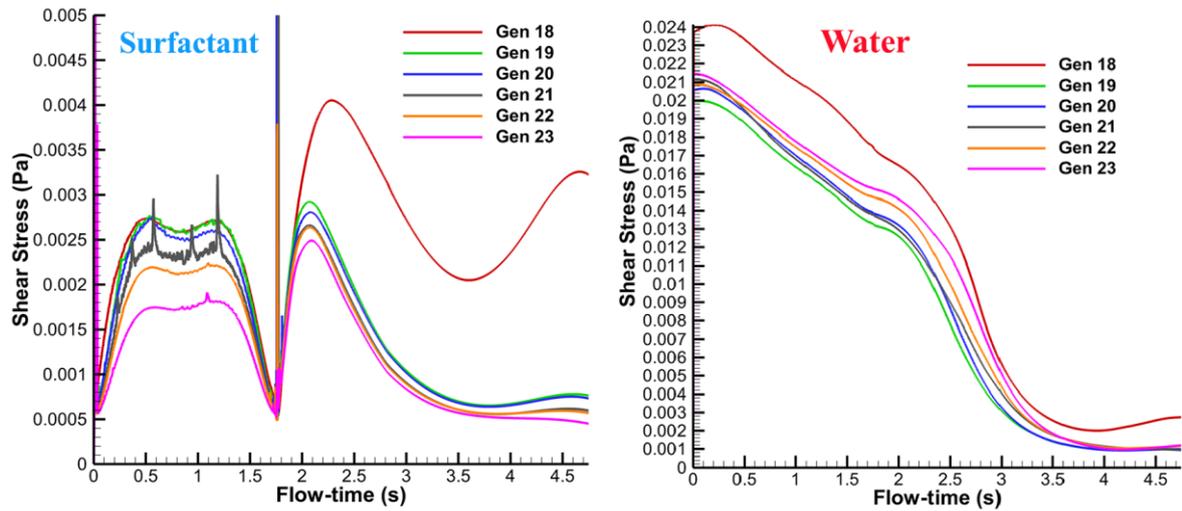

Figure 10: Area-weighted average of shear stress (Pa) on each generation surface for the surfactant case (left) and water case (right)

Figure 11 compares alveolar wall shear stress between the water and surfactant cases at different times during the inhalation phase. At all times, the water shear stress values are higher than those of the surfactant. At time instances 0.01s, 0.5s, 1, and 1.5s, water experiences shear stress at around 0.3 Pa, whereas surfactant experiences shear stress at around 0.002 Pa. In the surfactant case, Gen 18 shows slightly higher shear stress values than the other generations because of the unsteady flow behavior as discussed previously. Fig. 12 shows the exhalation phase at three instances, 2s, 3s, and 4s. Similar to the inhalation phase, water shear stresses are higher than the surfactant shear stresses. At 2s, the water and surfactant shear stresses are higher than those at 3s and 4s. This is because, at the beginning of exhalation at 2s, flow recirculates inside the alveoli, reversing the flow direction and creating large vortices and thus large shear stress values.

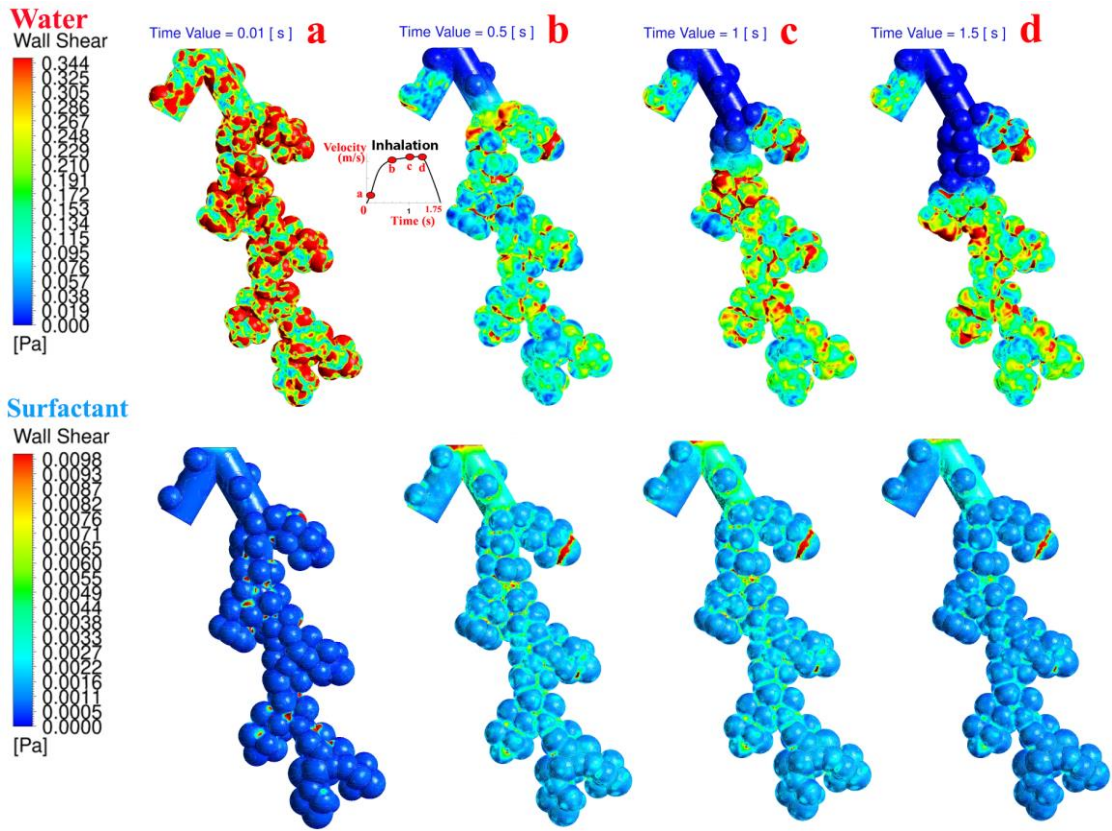

Figure 11: Illustration of the surface shear stress contours for the water (above) and surfactant (below) cases during the inhalation phase

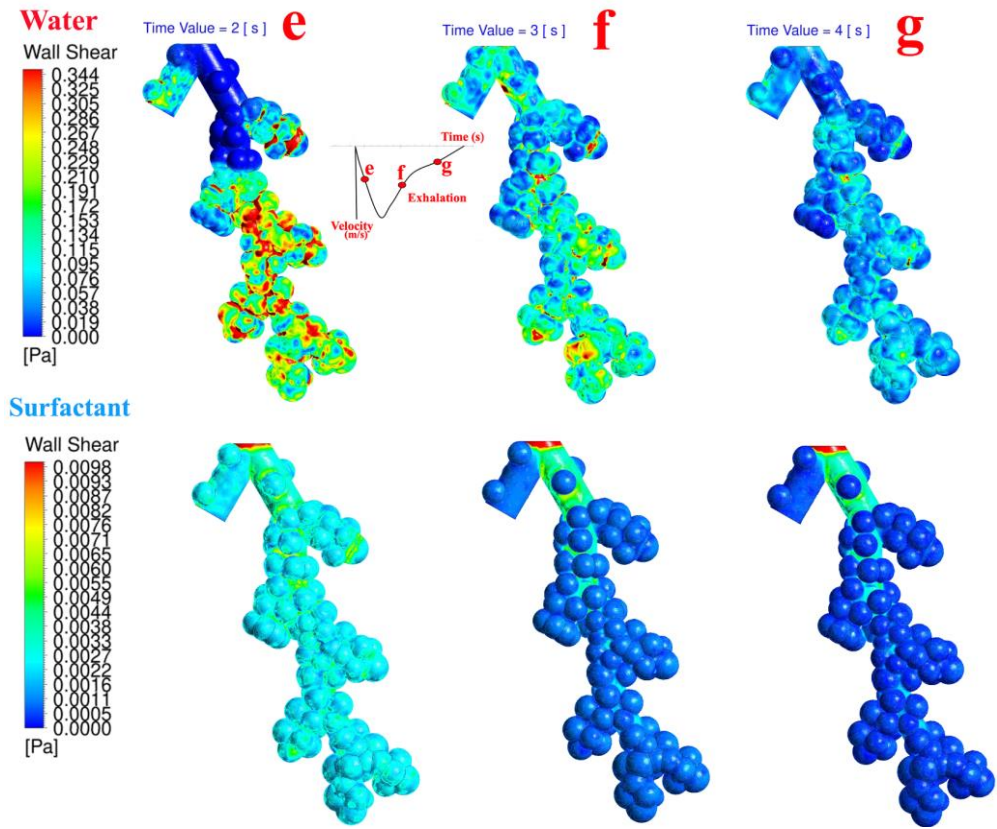

Figure 12: Three instantaneous snapshots of the water (above) and surfactant (below) wall shear stress during the exhalation phase

## 5. Conclusion:

Physiologically accurate acinar models with different patched liquids on their surfaces have been compared to assess alveolar and airflow alterations. The air-liquid surface tension has been shown to affect the alveolar surface mechanics, airflow velocities, and patterns. Specifically, according to Laplace's law, the high air-water surface tension increases the air pressure inside the alveoli, generating unsteady flow characteristics such as vortices and higher velocity, vorticity, and wall shear stresses. Consequently, these phenomena disrupt smooth normal breathing and may aggravate alveolar injuries if shear forces are elevated. Pulmonary surfactant prevents these from occurring by reducing the alveolar surface tension and thus improving general respiratory mechanics. Unfortunately, alveoli collapse and several other traumatic consequences of this unsteady flow usually develop in pre-born babies and ARDS patients with surfactant deficiency. This highlights the importance of administering exogenous pulmonary surfactants into surfactant-deficient lungs. Limitations of this study include an idealised acinar model with hemispherical alveoli and perfectly cylindrical ducts, symmetric alveolar airways, and rigid alveolar walls throughout the breathing cycle.

Further simulations of high surface tension effects will be performed with moving alveolar walls to obtain exact correlations between surface tension, vorticity, and shear stress. In addition, there is an apparent lack of substantial investigation into the particle kinematics near the alveolar surfaces. Future studies will address this issue by examining the inhalation of micron-sized particles and nanoparticles and the particle velocity magnitudes at alveolar surfaces under different breathing conditions and with the effects of high and low surface tension.

**Declaration of competing interest**

The authors declare no conflict of interest.

**Acknowledgement**

The authors would like to acknowledge the computing facility at University of Technology Sydney (UTS). This research did not receive any specific grant from funding agencies in the public, commercial, or not-for-profit sectors.